# Detecting Abnormal Profiles in Collaborative Filtering Recommender Systems

Zhihai Yang [1]

[1] Ministry of Education Key Lab for Intelligent Networks and Network Security, Xi'an Jiaotong University, Xi'an, 710049, China
E-mail: zhyang_xjtu@sina.com

*Abstract*—Personalization collaborative filtering recommender systems (CFRSs) are the crucial components of popular e-commerce services. In practice, CFRSs are also particularly vulnerable to "shilling" attacks or "profile injection" attacks due to their openness. The attackers can carefully inject chosen attack profiles into CFRSs in order to bias the recommendation results to their benefits. To reduce this risk, various detection techniques have been proposed to detect such attacks, which use diverse features extracted from user profiles. However, relying on limited features to improve the detection performance is difficult seemingly, since the existing features can not fully characterize the attack profiles and genuine profiles. In this paper, we propose a novel detection method to make recommender systems resistant to the "shilling" attacks or "profile injection" attacks. The existing features can be briefly summarized as two aspects including rating behavior based and item distribution based. We firstly formulate the problem as finding a mapping model between rating behavior and item distribution by exploiting the least-squares approximate solution. Based on the trained model, we design a detector by employing a regressor to detect such attacks. Extensive experiments on both the MovieLens-100K and MovieLens-ml-latest-small datasets examine the effectiveness of our proposed detection method. Experimental results were included to validate the outperformance of our approach in comparison with benchmarked method including KNN.

## 1. INTRODUCTION

To overcome the phenomenon of information overload, recommender systems (RSs) are essential tools to guide the customers to the information they are seeking for without wasting time navigating irrelevant information. Personalization RSs help users to select the items that they might like. They use information about user profiles to predict the relevance or utility of a particular product. Collaborative filtering is one of the most successful recommendation algorithm and used in different applications. Examples of collaborative filtering recommender systems (CFRSs) including Amazon and Ringo [1], [5], [8], [11]. However, CFRSs are prone to manipulation from producers or attackers. The phenomenon is often called "shilling" attacks or "profile injection" attacks. In such attacks, the attackers deliberately insert attack profiles into genuine user profiles to bias the predict results to their benefits, which can decrease the trustworthiness of recommendation. The attack profiles indicate the attacker's intention that he wishes a particular item can be rated with the highest score (called push attack) or the lowest score (called nuke attack). There are several reports about such attacks on Amazon or eBay. These events are that of a well-known company injecting malicious profiles into systems to promote its films [15], [17]. Therefore, constructing an effective method to defend the attackers and remove them from the CFRSs is crucial.

An interesting question from this case is how to systematically and dynamically detect the "shilling" attacks by using existing limited features extracted from a large number of user profiles. To answer this question is very important, since detecting "shilling" attacks usually carry interesting information of potential problems which requires real-time attention and should be detected and dealt with at the early stage. However, existing detection methods showed an acceptable level of performance in a few of attack models, which based on limited features [1], [5], [6], [11], [12], [14]. In addition, how to improve the detection performance when attack size [1] and filler size [2] are small is also a tricky problem [1], [12]. In practice, relying on the limited features to improve the detection performance is difficult seemingly. Hence, we either explore new and effective features to characterize user profiles or employ an advanced detection approach to improve the detection performance.

---

[1] The ratio between the number of attackers and genuine users.

[2] The ratio between the number of items rated by user $u$ and the number of entire items in the recommender systems.

As mentioned above, we assume that there is a correspondence relationship between the existing features which can be roughly divided into two categories (such as rating behavior based and item distribution based) and it can be used to discriminate between attack profiles and genuine profiles. Since attackers and genuine users may have different rating intentions on the rated items to generate different rating behaviors. Take bandwagon attack for example, the attackers select popular items and totally rate the highest or lowest score (in push or nuke attack), which may show an abnormal rating behavior. Comparing with attackers and genuine users, the genuine users may not totally rate the highest score on the popular items. As is known, popular items are not mean the best items. Based on the hypothesis, we try to find a mapping model to characterize this relationship. In this paper, we propose a novel detection method to make recommender systems resistant to "shilling" attacks. The objective of our proposed method is to be well used the extracted features and construct a mapping model for detecting "shilling" attacks. To achieve this objective, the key challenge is how to construct a relationship in the extracted features. To this end, we take both the features of rating behavior and item distribution into consideration when designing the mapping model. By adopting the least-squares approximate solution, we train the mapping model and design a detector to detect such attacks. In addition, the effectiveness of our proposed approach is validated and benchmarked methods are briefly discussed. Experimental results show that our approach performs well in comparison with the benchmarked methods.

The main purposes and major contributions of our paper are summarized as follows:

- We exploit 8 features to characterize the item distribution and employ 12 features to characterize the rating behavior for constructing a corresponding relationship between ratings and items.

- We propose a novel detection approach to detect "shilling" attacks based on the presented features, which adopts the relationship between rating behavior and item distribution to construct a mapping model.

- Our method is completely data-driven. We conduct experiments on both the MovieLens-100K and MovieLens-ml-latest-small datasets and compare the performance of our proposed method with benchmarked method including KNN.

The remaining parts of this paper are organized as follows: Section 2 reviews the related work on "shilling" attack detection in collaborative filtering recommender systems. Section 3 introduce attack profiles and attack models. The proposed detection method is introduced in Section 4. Experimental results are presented and discussed in Section 5. Finally, we briefly summarize our work and discuss the limitations of this paper and our future work.

## 2. RELATED WORK

Collaborative filtering recommender systems (CFRSs) are known to be vulnerable to "shilling" attacks or "profile injection" attacks. Shilling attack is a method by constructing attack profiles based on knowledge of the CFRSs to achieve attacker's intent. Published studies of attack detection in CFRSs have investigated different detection methods. Burke et al. [1] presented several features extracted from user profiles for their utility in attack detection. They exploited kNN classifier as their detection method. However, it was unsuccessful to detect attacks when filler size is small. Williams et al. [21], [22] tried to extract features from user profiles and utilized them to detect such attacks. But, they also suffered from low detection accuracy and some genuine profiles are misclassified as attack profiles. Then, He et al. [23] incorporated the rough set theory into detect shilling attacks by means of taking features of user profiles as the condition attributes of the decision table. But, their method also faced with the low overall classification rate in some cases, especially for bandwagon attack. After that, Wu et al. [18] proposed a hybrid detection algorithm to defend shilling attacks, which combines the naive Bayesian classifiers and augmented expectation maximization base on several selected metrics. Unfortunately, their approach also suffered from low F-measure [25] when filler size is small. In addition, Zhang et al. [24] proposed the idea of ensemble learning for improving predictive capability in attack detection. They constructed the base-classifiers (or weaker learner) with the Support Vector Machine (SVM) approach and then integrate them to generate a high predictive ability learner for detection. Their proposed method exhibited better performance than some benchmarked methods. Nevertheless, it still suffered from low precision especially when the attack size is small. And then, the same authors [14] also introduced an online method, HHT-SVM, to detect profile injection attacks by combining Hilbert-Huang transform (HHT) and support vector machine (SVM). They created rating series for each user profile based on the novelty and popularity of items in order to provide basic data for feature extraction. The precision of their method shown better than the benchmarked methods, but the precision significantly decreased with the filler size increased. Frankly speaking, previous studies showed that the detection results of "shilling" attacks are dissatisfactory and leave much to be desired, especially when the filler size or attack size is small. On the one hand, the existing features are limited so that it is difficult to fully characterize the attack profiles and genuine profiles. On the other hand, the detection performance of these detection methods is limited. To get a better use the existing features, we roughly divide these features into two categories including rating behavior and item distribution and find a trained mapping model to construct a relationship between the two categories for detecting the "shilling" attacks, since attackers and genuine users have different rating intentions on different items.

## 3. ATTACK PROFILES AND ATTACK MODELS

The attackers have different attack intents which bias the recommendation results to their benefits. In the literature, the shilling attacks are classified into two ways: push attack and nuke attack [1], [21], [22]. In push attack, attackers promote the target items by rating the highest score, whereas in nuke attack, attackers demote the target items by rating the lowest score. In order to push or nuke a target item, the attacker should be clearly known the form of attack profiles. The general form of attack profiles is shown in Table 1. The details of the four sets of items are described as follows:

$I_S$: The set of selected items with specified rating by the function $\sigma(i_k^S)$ [10];

$I_F$: A set of filler items, received randomly items with random assigned rating by the function $\rho(i_l^F)$;

$I_N$: A set of items with no ratings;

$I_T$: A set of target items with singleton or multiple items, called single target attack or multiple target attack. The rating is $\gamma(i_j^T)$, generally rated the maximum or minimum value in the entire profiles.

In this paper, we introduce 8 attack models to generate attack profiles. The presented attack models and corresponding explanations are listed in Table 2. The details of these attack models are described as follows:

1) AOP attack: A simple and effective strategy to obfuscate the Average attack is to choose filler items with equal probability from the top x% of most popular items rather than from the entire collection of items [26].

2) Random attack: $I_S = \emptyset$ and $\rho(i) \sim N(\bar{r}, \bar{\sigma}^2)$ [6].

3) Average attack: $I_S = \emptyset$ and $\rho(i) \sim N(\bar{r}_i, \bar{\sigma}_i^2)$ [6].

4) Bandwagon (average): $I_S$ contains a set of popular items, $\sigma(i) = r_{max}/r_{min}$ (push/nuke) and $\rho(i) \sim N(\bar{r}_i, \bar{\sigma}_i^2)$ [7].

5) Bandwagon (random): $I_S$ contains a set of popular items, $\sigma(i) = r_{max}/r_{min}$ (push/nuke) and $\rho(i) \sim N(\bar{r}, \bar{\sigma}^2)$ [7].

6) Segment attack: $I_S$ contains a set of segmented items, $\sigma(i) = r_{max}/r_{min}$ (push/nuke) and $\rho(i) = r_{min}/r_{max}$ (push/nuke) [14].

7) Reverse bandwagon attack: $I_S$ contains a set of unpopular items, $\sigma(i) = r_{min}/r_{max}$ (push/nuke) and $\rho(i) \sim N(\bar{r}, \bar{\sigma}^2)$ [14].

8) Love/Hate attack: $I_S = \emptyset$ and $\rho(i) = r_{min}/r_{max}$ (push/nuke) [14].

TABLE I. GENERAL FORM OF ATTACK PROFILES.

| $I_T$ | | | | $I_S$ | | | $I_F$ | | | $I_N$ | | |
|---|---|---|---|---|---|---|---|---|---|---|---|---|
| $i_1^T$ | ... | $i_j^T$ | $i_1^S$ | ... | $i_k^S$ | $i_1^F$ | ... | $i_l^F$ | $i_1^N$ | ... | $i_v^N$ |
| $\gamma(i_1^T)$ | ... | $\gamma(i_j^T)$ | $\sigma(i_1^S)$ | ... | $\sigma(i_k^S)$ | $\rho(i_1^F)$ | ... | $\rho(i_l^F)$ | null | ... | null |

TABLE II. ATTACK MODELS SUMMARY.

| Attack Model | $I_S$ | | $I_F$ | | $I_N$ | $I_T$ (push/nuke) |
|---|---|---|---|---|---|---|
| | *Items* | *Rating* | *Items* | *Rating* | | |
| AOP | null | | x-% popular items, ratings set with normal dist around item mean. | | null | $r_{max}/r_{min}$ |
| Random | null | | randomly chosen | system mean | null | $r_{max}/r_{min}$ |
| Average | null | | randomly chosen | item mean | null | $r_{max}/r_{min}$ |
| Bandwago (average) | popular items | $r_{max}/r_{min}$ | randomly chosen | item mean | null | $r_{max}/r_{min}$ |
| Bandwagon(random) | popular items | $r_{max}/r_{min}$ | randomly chosen | system mean | null | $r_{max}/r_{min}$ |
| Segment | segmented items | $r_{max}/r_{min}$ | randomly chosen | $r_{min}/r_{max}$ | null | $r_{max}/r_{min}$ |
| Reverse Bandwagon | unpopular items | $r_{min}/r_{max}$ | randomly chosen | system mean | null | $r_{max}/r_{min}$ |
| Love/Hate | Null | | randomly chosen | $r_{min}/r_{max}$ | null | $r_{max}/r_{min}$ |

## 4. OUR PROPOSED APPROACH

In this section, we firstly introduce the problem formulation and then describe the framework of the proposed method. Finally, we describe our detection method.

### 4.1. Problem Formulation

#### A. Model Structure

From the relationship between rating and item of view, we explore a model structure to reveal the dependency relationship between rating behavior and item distribution. So, we can write loosely

$$i = F(r) + \epsilon, \qquad (1)$$

where i is a vector of output, namely the attributes of item distribution, r is a vector of input, namely the attributes of rating behavior, and $\epsilon$ is some noise that is a consequence of unaccounted measure attributes and turbulence. The nonlinear map F(r) describes the dependency relationship between rating behavior and item distribution. In addition, the components of the input vector r including twelve rating attributes (as shown in Table 4) including RDMA, WDMA, etc. The components of the output vector i including eight attributes of item distribution (as shown in Table 5), such as FSTI, FSPI, etc. The goal is to solve for the function F and use this model to detect "shilling" attacks. For rating attributes, we use the existing rating metrics to measure rating patterns of "shilling" attacks. The presented metrics (see Table 4) provide a useful insight into making a distinction between attack profiles and genuine profiles [1], [2], [3], [12], [16], [20]. In this paper, we incorporate the twelve rating attributes into our model structure. In the early stage, we calculate all rating attributes for each user and normalize the results to construct our input vector.

TABLE III. NOTATIONS AND THEIR DESCRIPTIONS.

| Notation | Description |
|---|---|
| $r_{u,i}$ | The rating assigned by user u to item i. [1] |
| $\bar{r}_i$ | The average rating value of item i. [1] |
| $NR_i$ | The set of users who have rated item i. [1] |
| $N_u$ | The number of items rated by user u. |
| $U_u$ | The set of items rated by user u. |
| $P_{u,T}$ | $P_{u,T} = \{i \in P_u, \text{such that } r_{u,i} = r_{max}\}$. [1] |
| $n_u$ | The length of profile u. |
| $\bar{n}$ | The average profile length in the system. |
| I | The set of the entire items in the recommender system. |

TABLE IV. RATING ATTRIBUTES.

| Attributes | Equations | Descriptions |
|---|---|---|
| RDMA | $RDMA_u = \dfrac{\sum_{i=0}^{N_u} \frac{|r_{u,i} - \bar{r}_i|}{NR_i}}{N_u}$ | Rating Deviation from Mean Agreement [1]. |
| WDMA | $WDMA_u = \dfrac{\sum_{i=0}^{N_u} \frac{|r_{u,i} - \bar{r}_i|}{NR_i^2}}{N_u}$ | Weighted Deviation from Mean Agreement [1]. |
| WDA | $WDA_u = \sum_{i=0}^{N_u} \dfrac{|r_{u,i} - \bar{r}_i|}{NR_i}$ | Weighted Degree of Agreement [1]. |
| FMD | $FMD_u = \dfrac{1}{|U_u|} \sum_{i=1}^{|U_u|} |r_{u,i} - \bar{r}_i|$ | Filler mean difference [12]. |
| MeanVar | $MeanVar_u = \dfrac{\sum_{j \in P_{u,F}} (r_{u,j} - \bar{r}_u)^2}{|P_{u,F}|}$ | Mean Variance [12]. |
| FMV | $FMV_u = \dfrac{1}{|U_u^F|} \sum_{I_i \in U_u^F} (r_{u,i} - \bar{r}_i)^2$ | Filler mean variance [12]. |
| FMTD | $FMTD_u = \left| \dfrac{\sum_{i \in P_{u,T}} r_{u,i}}{|P_{u,T}|} - \dfrac{\sum_{k \in P_{u,F}} r_{u,k}}{|P_{u,F}|} \right|$ | Filler Mean Target Difference [12]. |
| TMF | $TMF_u = \max_{j \in P_T} F_j$ | Target Model Focus attribute [0] [12]. |
| LengthVar | $LengthVar_u = \dfrac{|n_u - \bar{n}|}{\sum_{k \in U} (n_k - \bar{n})^2}$ | Length Variance [1]. |
| Entropy | $Entropy_u = -\sum_{r=1}^{r_{max}} \dfrac{n_{r_u}}{\sum_{i=1}^{r_{max}} n_{i_u}} \log_2 \dfrac{n_{r_u}}{\sum_{i=1}^{r_{max}} n_{i_u}}$ | Entropy of user rating vector [12]. |
| FAC | $FAC_u = \dfrac{\sum_i (r_{u,i} - \bar{r}_i)}{\sqrt{\sum_i (r_{u,i} - \bar{r}_i)^2}}$ | Filler average correlation [12]. |
| UNRAP | $H_v(u) = \dfrac{\sum_{i \in I} (r_{ui} - r_{Ui} - r_{uI} + r_{UI})^2}{\sum_{i \in I} (r_{ui} - r_{uI})^2}$ | Unsupervised Retrieval of Attack Profiles [1]. |

TABLE V. ITEM ATTRIBUTES.

| Attributes | Equations | Descriptions |
|---|---|---|
| Filler Size with Total Items (FSTI) | $FSTI_u = \frac{\sum_{i=1}^{|I|} O(r_{u,i})}{|I|}$ | The ratio between the number of items rated by user u and the number of entire items in the recommender system [14]. |
| Filler Size with Popular Items (FSPI) | $FSPI_u = \frac{\sum_{i=1}^{K} O(r_{u,i})}{K}$ | The ratio between the number of popular items rated by user u and the number of entire popular items in the recommender system [14]. |
| Filler Size with Popular Items in Itself (FSPII) | $FSPII_u = \frac{\sum_{i=1}^{K} O(r_{u,i})}{\sum_{j=1}^{|I|} O(r_{u,j})}$ | The ratio between the number of popular items rated by user u and the number of entire items rated by user u [14]. |
| Filler Size with Unpopular Items (FSUI) | $FSUI_u = \frac{\sum_{i=1}^{|I|} O(r_{u,i})}{U}$ | The ratio between the number of unpopular items rated by user u and the number of entire unpopular items in the recommender system [14]. |
| Filler Size with Unpopular Items in Itself (FSUII) | $FSUII_u = \frac{\sum_{i=1}^{|I|} O(r_{u,i})}{\sum_{k=1}^{|I|} O(r_{u,k})}$ | The ratio between the number of unpopular items rated by user u and the number of entire items rated by user u [14]. |
| Filler size with maximum rating in total items (FSMAXRTI) | $FSMAXRTI_u = \frac{\sum_{i=1}^{|I|} O(r_{u,i} = r_{max})}{|I|}$ | The ratio between the number of items rated by user u with maximum score and the number of entire items in the recommender system. |
| Filler size with minimum rating in total items (FSMINRTI) | $FSMINRTI_u = \frac{\sum_{i=1}^{|I|} O(r_{u,i} = r_{min})}{|I|}$ | The ratio between the number of items rated by user u with minimum score and the number of entire items in the recommender system. |
| Filler size with average rating in total items (FSARTI) | $FSARTI_u = \frac{\sum_{i=1}^{|I|} O(r_{u,i} = r_{ave})}{|I|}$ | The ratio between the number of items rated by user u with average score and the number of entire items in the recommender system. |

To explore the internal relevance between rating and item, we firstly observe the distribution of filler items or selected items in user profiles. We conduct a list of attack experiments by using 8 attack models (as shown in Table 2). One observation is that the selected items contain popular and unpopular items. In addition, from the items that rated maximum rating (for push attack) or minimum rating (for nuke attack) of view, there are three ratings (maximum score 5, minimum score 1 and average score) that rate on the selected items or filler items. Based on the above mentioned, we propose 8 metrics to measure the distribution of items (such as FSTI, FSPI, etc. [14]) as shown in Table 5 (we give the descriptions of notations used in this paper to facilitate the discussions in Table 3).

*B. Problem Statement*

The problem consists of two phases: the first is to identify the model F; the second is to use the trained model to detect shilling attacks. These correspond to the two major blocks in Figure 1. We firstly describe our data of user profiles more formally. For a user profile, we have a number of items that rated by the user and his ratings. A user might rate at least one item. These items come from different item distribution, such as popular, unpopular, etc. In addition, the user has corresponding rating on each item, which constructs rating behavior of the user. So, the data can be represented as

$$\{R_u, I_u\}_{u=1}^N, \qquad (2)$$

where N is the number of users, R and I are the input and output vector, respectively. The input-output pairs are independent and obey Eq. (1).

The first problem is to identify the nonlinear map F from the training set. We approach this as a linear regression problem and replace Eq. (1) with

$$I = M\Psi(R) + e, \qquad (3)$$

where M is a matrix that contains the model parameters, $\Psi(R)$ is a regressor, and e denotes the noise which combines $\epsilon$ in Eq. (1) and the error of the regression fit. In what follows, we will discuss the choice of regressors.

The second problem is to use the trained model M to detect anomalies. We consider a modification of Eq. (3) and write it as

$$I = M\Psi(R) + e + f, \qquad (4)$$

where $f$ denotes the impact of the shilling attacks on the output I. Assuming that R, I and M are known in Eq. (4), we want to determine if $f$ is nonzero. Then we call this problem the anomaly detection problem:

TABLE VI. THE PREDICTIVE POWER OF THE LINEAR REGRESSOR AND QUADRATIC REGRESSOR IN DIFFERENT ATTACK MODELS.

| Attack Model | Predictive Power | |
|---|---|---|
| | *Linear Regressor* | *Quadratic Regressor* |
| AOP | 0.765 | 0.808 |
| Random | 0.778 | 0.812 |
| Average | 0.785 | 0.810 |
| Bandwago (average) | 0.813 | 0.845 |
| Bandwagon(random) | 0.821 | 0.842 |
| Segment | 0.807 | 0.839 |
| Reverse Bandwagon | 0.793 | 0.831 |
| Love/Hate | 0.805 | 0.827 |

problem: we want to identify when there exists attack profile in an entire user profiles given by (2). In addition, we will describe how to detect shilling attack in the following section.

To be brief, the problem can be summarized as follows:

1) Find a model that holds across a range of user profiles;
2) Use the trained model to detect shilling attacks;

*C. Choice of Regressors*

To address our problem, we can use different functions $\Psi$ to map the inputs and outputs. A simple one might be the linear regressor

$$\Psi_{\text{linear}}(R) = [R, 1]^T. \qquad (5)$$

where R is the input vector. By using this regressor, the trained model M would describe a linear relationship between the input R and the output I. In this paper, we also use an alternative quadratic regressor that is constructed as follows:

$$\Psi_{\text{quadratic}}(x) = \left[\{R_j R_k\}_{j \le k}, R, 1\right]^T \qquad (6)$$

where $\{R_j R_k\}_{j \le k}$ is the set of all distinct rating behaviors of the elements of the input vector R. The subscript refers to the elements of R as opposed to a distinct rating behavior. Similarly, the trained model M would describe a quadratic relationship between the input R and output I [19]. Of course, we can choose other high-dimensional regressors in our work. It's worth noting that the choice of regressors affects the performance and interpretation of the regression model, but not how the model is constructed.

In our work, we conduct a list of experiments on different attack models to choose a better regressor. As shown in Table 6, we report the predictive power [19] (as shown in Eq. (17)) of the trained model for both linear regressor and quadratic regressor in different attack strategies. In addition, we fix 12.5% attack size and 5.2% filler size in each attack model. The results may indicate that quadratic regressor has better performance than linear regressor as illustrated in Table 6. Therefore, we choose the quadratic regressor as the regressor in our work.

*D. Model Identification with Large Amounts of Data*

Model identification gives us an empirical model which can be used for anomaly detection. In this part, we illustrate how to train a model M from a large dataset. By using data without heavy turbulence (so that e is small) and without any impact of the attacks on the output I (such that f=0), we can find a model from the data. The solution can be constructed by finding the least squares approximate solution for M by solving

$$\text{minimize } \sum_{u=1}^{N} \|I_u - M\Psi_u\|^2 + \lambda \|M\|_F^2, \qquad (7)$$

where $I_u$ and $\Psi_u$ denote the output vector and the regressor, respectively. $\|\cdot\|_F$ denotes the Frobenius norm of a matrix. N denotes the number of users. The parameter $\lambda$ is chosen to prevent model over-fitting and keep the coefficients in the model "small". In addition, we call the least-squares model obtained this way the linear model if we use a linear regressor for the model. In the same way, a different regressor would be described similarly (such as a quadratic model) [19].

To compute quickly the solution Eq. (7) and overcome a distributed fashion for a large dataset that do not fit into memory or store in large historical datasets, we take the derivative with respect to the matrix M in Eq. (7) and set the result equal to zero, and then we obtain the normal equations that provide the solution to Eq. (7) as follows:

$$\frac{\partial}{\partial M}(\sum_{u=1}^{N}\|I_u - M\Psi_u\|^2 + \lambda\|M\|_F^2) = 0,$$

$$(\lambda I + \sum_{u=1}^{N} \Psi_u\Psi_u^T)M^T = \sum_{u=1}^{N} \Psi_u I_u^T. \tag{8}$$

After the derivation of the formula, we can construct our model recursively and also easily handle large datasets by updating the two matrices containing these running sums,

$$\sum_{u=1}^{N} \Psi_u\Psi_u^T \text{ and } \sum_{u=1}^{N} \Psi_u I_u^T. \tag{9}$$

In the end, we calculate the empirical covariance for the residuals as a combination of the running sums

$$C = \frac{1}{N-1} \sum_{u=1}^{N}(I_u - M\Psi_u)(I_u - M\Psi_u)^T$$

$$= \frac{1}{N-1}(\sum_{u=1}^{N}(I_u I_u^T) - 2M\sum_{u=1}^{N}(\Psi_u I_u^T) + M\sum_{u=1}^{N}(\Psi_u \Psi_u^T) M^T) \tag{10}$$

where N is the number of users, and M is our trained model which can be solved by adding the regularization $\lambda I$. Note that we have already computed the matrices, $\sum_{u=1}^{N}(\Psi_u I_u^T)$ and $\sum_{u=1}^{N}(\Psi_u \Psi_u^T)$. We also calculate the matrix $\sum_{u=1}^{N}(I_u I_u^T)$ in a similar manner.

*4.2. The Framework of the Proposed Method*

As shown in Figure 1, our proposed algorithm consists of two stages: the stage of normalizing data and the stage of detection. At the stage of normalizing data, we are firstly given the user profile data, including training set and test set. To capture the essential information for our proposed method in the early phase, we construct the vectors of rating attributes and item attributes (see subsection (4.1)). And then we normalize our data and apply the regressor to the normalized data. At the stage of detection, the training set in the normalized data is used to train the model of classification and the empirical covariance for the purpose of detection. In addition, at the stage of detection, we train the model M by solving Equations (7)-(11). And then, we calculate the empirical covariance for the residuals as a combination of the running sums C by solving Eq. (10). By using the trained model M and the empirical covariance C, we firstly calculate the residual part (as shown Eq. (11)) in test dataset. To detect anomaly profiles, we use a detector D to decide whether a user profile is anomalous [19]. In our work, we incorporate the Mahalanobis norm into our detection method. It can be summarized as a detector (as shown in Eq. (12)), which is a function of the individual residuals. The details of detection will be discussed in the following subsection.

Let *I* and *Ψ* denotes the output vector and regressor, respectively. The proposed data normalization algorithm is described in algorithm 1. In this algorithm, step 1 and step 2 create respectively the vector of rating and item attributes for each user u. These attributes are described in Tables 3 and 4. Step 3 normalize the input data by using Eq. (16) and apply the regressor *Ψ* to the normalized data. Step 4 raises a linear regression problem which is the basis for our proposed method.

| **Algorithm 1:** Data normalization algorithm for user profiles |
|---|
| **Input:** Rating matrix; |
| **Output:** Regressor *Ψ*; Vector *I*; |
| **Step 1:** Create the vector of rating attributes $R_u$ for each user $u$ by using Table 4; |
| **Step 2:** Create the vector of item attributes $I_u$ for each user $u$ by using Tables 5; |
| **Step 3:** Normalize data by using Eq. (15) and apply regressor *Ψ* by using Eq. (5) or (6); |
| **Step 4:** Based on the relationship between rating and item, constructing linear regression problem (see Eq. (3)) by using Eq. (1); |
| **Step 5:** Generate and return *Ψ* and *I*. |

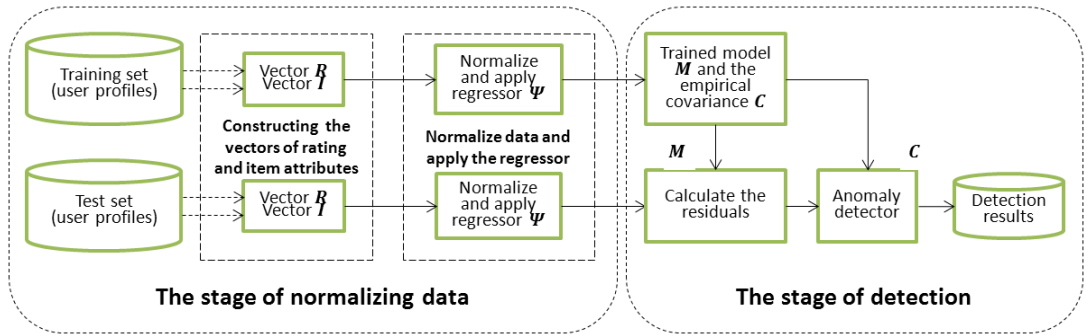

Fig. 1. The framework of our proposed method which consists of two stages: the stage of normalizing data and the stage of detection.

*4.3. Anomaly Detection*

To design detection method, we firstly use the trained model M and calculate the residual R in the test sets. The formula of the residual can be represented as

$$Res_u = I_u - M\Psi_u. \quad (11)$$

where u is the users in the test sets. The motivation for using this method inspired from Eric Chu et al. [19].

Based on the given residuals Res and the empirical covariance C, we could set a threshold T by using the norm of residual for a user profile. This is similar to calculating the Hotelling $H^2$ statistic for all user profiles. The detector labels every user as attacker or not by comparing the Mahalanobis norm of the residual to a threshold T. In addition, we use the empirical covariance C to calculate the Mahalanobis norm. The detector can be represented as

$$D(Res_u) = \begin{cases} 1 & \text{if } \left\|Res_u C^{-\frac{1}{2}}\right\|^2 \geq T \\ 0 & \text{otherwise,} \end{cases} \quad (12)$$

where C is the empirical covariance, and T is the chosen threshold.

Let ***D*** denotes the detection results. The proposed detection method is described in algorithm 2. In the algorithm 2, step 1 is the process of training model. Step 2 calculates the empirical covariance ***C*** by using the trained model ***M***. Step 3 calculates the residual by using Eq. (11), which is prepared for the detector. Algorithm 2 can detect shilling attacks once the detector is generated.

---
**Algorithm 2:** Detection algorithm
**Input:** Regressor $\boldsymbol{\Psi}_{train}$ and $\boldsymbol{\Psi}_{test}$; Vectors $\boldsymbol{I}_{train}$ and $\boldsymbol{I}_{test}$;
**Output:** The detected results ***D***;
**Step 1:** Training model ***M*** by using $\boldsymbol{\Psi}_{train}$, $\boldsymbol{I}_{train}$ and Equations (8)-(9);
**Step 2:** Calculating the empirical covariance ***C*** by using Eq. (10);
**Step 3:** Calculating the residual by using ***M***, $\boldsymbol{\Psi}_{test}$, $\boldsymbol{I}_{test}$ and Eq. (11) for each user $u$ in test sets;
**Step 4:** Generate and return ***D*** by using Eq. (12) and ***C*** to classify each user $u$ in the test set;

---

## 5. EXPERIMENTS AND ANALYSIS

In this section, we firstly introduce the experimental data and settings on two real-world datasets. Then we describe attack profiles and evaluation metrics. Finally, we conduct a list of experiments on the two datasets and validate the effectiveness of our proposed method.

*5.1. Experimental Data and Settings*

In our experiments, we use both the MovieLens-100K [3] and MovieLens-ml-latest-small [4] datasets as the datasets describing the behaviors of genuine users in recommender systems. MovieLens datasets were collected by the GroupLens Research Project at the University of Minnesota. The two datasets are the most popular datasets used by researchers and developers in the field of collaborative filtering and attack detection in recommender systems. The MovieLens-100K dataset consists of 100,000 ratings on 1682 movies by 943 raters and each rater had to rate at least 20 movies. All ratings are in the form of integral values between minimum value 1 and maximum value 5. The minimum score means the rater distastes the movie, while the maximum score means the rater enjoyed the movie. According to the information derived from MovieLens website, the sparse ratio [5] of the rating data approximates to 93.7% and the average rating of all users is around 3.53. Besides, the Average Number of Items Rated (ANIR) by each user is approximately 7%. For the MovieLens-ml-latest-small dataset, it describes 5-star rating and free-text tagging activity from a movie recommendation service. It contains 100023 ratings and 2488 tag applications across 8570 movies. These data were created by 706 users between April 02, 1996 and March 30, 2015. This dataset was generated on April 01, 2015. Users were selected at random for inclusion. All selected users had rated at least 20 movies.

In our attack experiment, attack profiles are created according to different attack models as shown in Table 2. The attack profiles indicate the attackers intention that he wishes a particular item can be rated the highest or the lowest. In this paper, we just detect the nuke attacks. Of course, our proposed approach can be used to detect the push attacks. Take MovieLens-100K dataset for example, we use a separate training and test set by partitioning the MovieLens-100K dataset in half. The first half was used to construct training dataset (470 genuine users) for training classifier. For each test the second half of the data was injected with attack profiles and then run through the classifier that had been built on the augmented first half. For each attack model, we generate samples of nuke attack profiles according to the corresponding attack model with different attack sizes {2.5%, 7.5%, 12.5%, 17.5%, 22.5%, 27.5%, 32.5%, 37.5%, 42.5%, 47.5} and filler sizes {1.3%, 3.2%, 5.2%, 7.1%, 9.1%, 11%, 13%}. To ensure the rationality of the results, the target item is randomly selected for each attack profile. In addition, we randomly select 10 movies as the popular movies which are rated by more than 200 users in the system. For reverse bandwagon attack, we randomly choose 10 movies as the selected movies which are rated by one user in the system. And for segment attack, we select 5 top popular items as the segmented items which are rated with the highest number of users.

---
[3] http://grouplens.org/datasets/movielens/
[4] http://grouplens.org/datasets/movielens/
[5] The ratio between the number of ratings and entire ratings in the rating matrix.

The training data was created by inserting a mix of attack profiles which combine diverse attack profiles (8 attack models) with different filler sizes {1.3%, 3.2%, 5.2%, 7.1%, 9.1%, 11%, 13%}. To balance the proportion between genuine profiles and attack profiles in the training set, we construct 60 attack profiles (attack size is 12.5%) for each of the mentioned attack models corresponding to the filler sizes. For test datasets, the second half of data was injected attack profiles with diverse filler sizes {1.3%, 3.2%, 5.2%, 7.1%, 9.1%, 11%, 13%} and attack sizes {2.5%, 7.5%, 12.5%, 17.5%, 22.5%, 27.5%, 32.5%, 37.5%, 42.5%, 47.5} for each attack model, respectively. Therefore, we have 560 (8*10*7) test datasets including 8 attack models, 10 different attack sizes and 7 different filler sizes. For the MovieLens-ml-latest-small dataset, we generate attack profiles in the same way. These processes are repeated 10 times and the average values of detection results are reported for the experiments. In addition, we should not forget that algorithms are strongly dependent on the characteristics of the dataset, so the results obtained in our study may not coincide with those obtained with data from other domains. All numerical studies are implemented using MATLAB R2012a on a personal computer with Intel(R) Core(TM) i7-4790 3.60GHz CPU, 16G memory and Microsoft windows 7 operating system.

To measure the effectiveness of the proposed detection methods, we use detection rate and false alarm rate in our paper.

$$\text{detection rate} = \frac{|D \cap A|}{|A|}, \tag{13}$$

$$\text{false alarm rate} = \frac{|D \cap G|}{|G|}, \tag{14}$$

where D is the set of the detected user profiles, A is the set of attacker profiles, and G is the set of genuine user profiles [9].

To avoid numerical problems and neutralize the effect of large variables, we normalized our data and apply the regressor $\Psi$ to the normalized data. Especially for the vector of rating attributes, the input vector consists of twelve attributes which have different measurement scales. We use the formula as follows:

$$\Psi_u = \Psi\left(2\left(\frac{R_u - R_{min}}{R_{max} - R_{min}}\right) - 1\right). \tag{15}$$

where $R_{min}$ and $R_{max}$ are the minimum and maximum value in the rating attributes (input), respectively. In addition, we also normalize the attributes of item distribution (output) in the same way (but do not apply the regressor). The normalization forces most of our raw data to lie in [-1, 1] and prevents large numbers from dominating [19].

To measure how good our trained model is, we want to know how of the output variation the model explains. We calculate the predictive power of the model by using following formula

$$pp = 1 - \frac{\sum_{u=1}^{N} \|I_u - M\Psi_u\|^2}{\sum_{u=1}^{N} \|I_u\|^2}, \tag{16}$$

Note that if $pp = 1$, then the numerator must have been 0. This only occurs if $M\Psi_u$ is a prefect predictor of $I_u$. If $pp = 0$, then we know that $M\Psi_u$ explains none of the variance in $I_u$ [19].

### 5.2. Experimental Results and Analysis

To validate the effectiveness of our proposed detection method in comparison with a benchmarked method KNN, we conduct a list of experiments on both MovieLens-100K and MovieLens-ml-latest-small datasets in each attack model with diverse filler sizes {1.3%, 3.2%, 5.2%, 7.1%, 9.1%, 11%, 13%} and attack sizes {2.5%, 7.5%, 12.5%, 17.5%, 22.5%, 27.5%, 32.5%, 37.5%, 42.5%, 47.5}. For KNN classifier, we utilize all 20 features extracted from user profiles including attack profiles and genuine profiles (as shown in Tables 4 and 5). Based on the training data described in subsection 5.1, KNN with k=9 was used to make a binary profile classifier with 0 if classified as authentic and 1 if classified as attack. Classification results and KNN classifier were created using Waka [6].

To compare the detection performance of our proposed method with the benchmarked method, we conduct a list of experiments with different attack sizes and filler sizes in two different attack models (average and bandwagon (random) attacks are taken for examples). As shown in Figure 2, by fixing 5.2% filler size, the curves of detection performance show a relatively stable trend and perform no change obviously with the attack size increased in the later stage (attack size > 22.5%) as illustrated in Figures 2(a) and 2(b). Notice that, the detection performance significantly outperforms KNN in both average and bandwagon (random) attacks. By fixing 7.5% attack size, the detection performance of our method also significantly outperforms KNN with the filler size increasing as illustrated in Figure 3. The results may indicate that our method is more effective to use the existing features than KNN. In some cases, more features used for a classifier may cause bad results (the curves of KNN as shown in Figure 3).

To compare the detection performance of our method in both nuke and push attacks, we plot the receiver operating curves (ROC) in Figure 4. Take bandwagon (random) attack for example, the curves show the dependence of the detection rate (Eq. (13)) and false alarm rate (Eq. (14)) on the threshold T in Eq. (12). The curves were empirically obtained using the test set data. In Figure 4, we can observe that the detection achieves 91.2% detection rate when the false alarm rate achieves 11.7% in nuke attack. For the push attack, the curve shows similarly phenomenon. In summary, the proposed method performs well by sacrificing false alarm rate to improve detection rate as illustrated in Figure 4. Our goal is to identify the effectiveness of our method in different attack intentions (nuke attack and push attack).

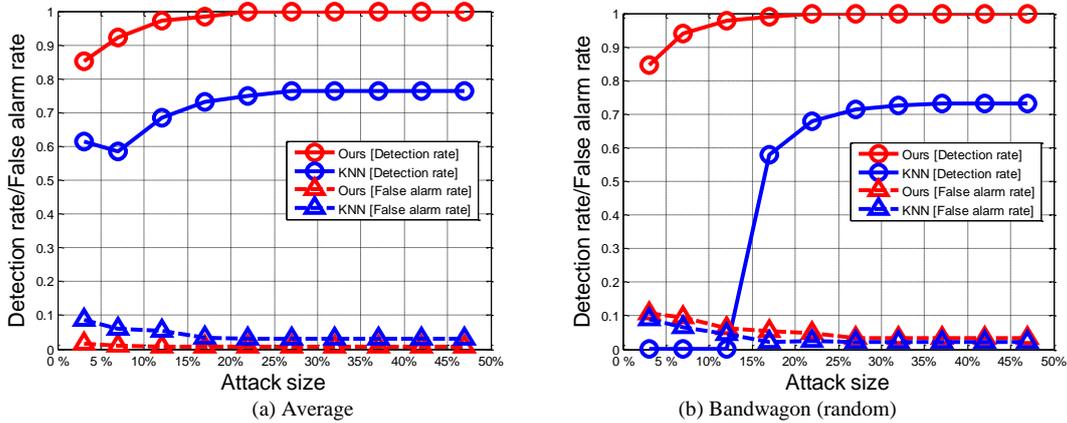

Fig. 2. The comparison of detection rate and false alarm rate when filler size 5.2% and attack size varies. (a) Average attack. (b) Bandwagon (random) attack.

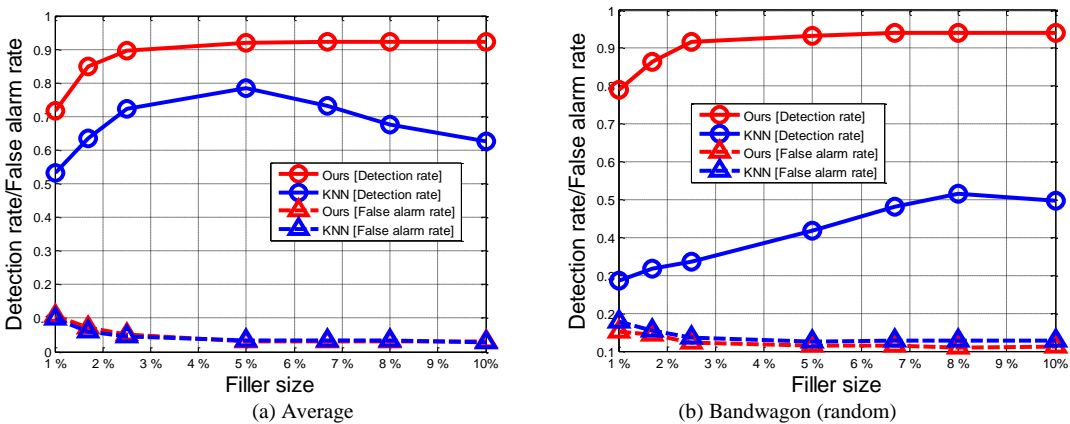

Fig. 3. The comparison of detection rate and false alarm rate when attack size is 7.5% and filler size varies. (a) Average attack. (b) Bandwagon (random) attack.

---

[6] http://www.cs.waikato.ac.nz/~ml/weka/

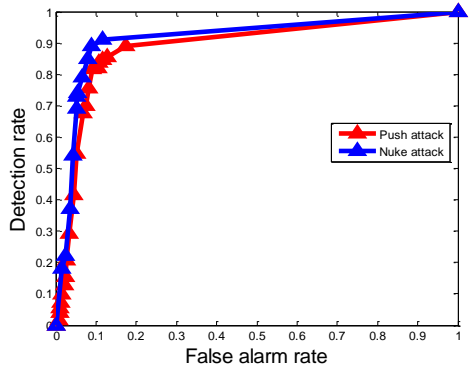

Fig. 4. ROC curves for the proposed method in nuke and push bandwagon (random) attack when attack size is 12.5% and filler size is 5.2%.

To further evaluate the effectiveness of our method, we also conduct a list of experiments by using other 6 attacks to show the performance curves of the proposed method just as Figures 5 and 6 illustrated. From these results, we can observe that our method can effectively detect all of attacks. Comparing with Figure 5 and Figure 6, the detection rate increased more obvious with attack size varies (see Figure 5) than filler size varies (see Figure 6). The second observation is that the detection rate ultimately achieves the highest value with the attack size increasing. Nevertheless, the detection rate shows low value when attack size is small. The results may indicate that it is difficult to fully solve the imbalanced classification by our method. In addition, the detection rate gradually increased and achieved a stable rate with the filler size increasing. These results may indicate that the presented features are also limited for our method to detect shilling attacks.

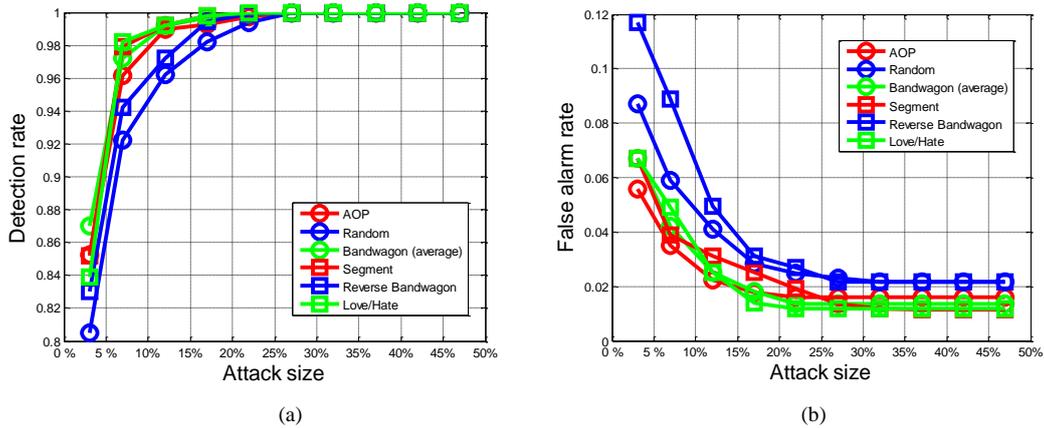

Fig. 5. The comparison of detection rate and false alarm rate in diverse attack models when filler size is 5.2% and attack size varies. (a) Detection rate. (b) False alarm rate.

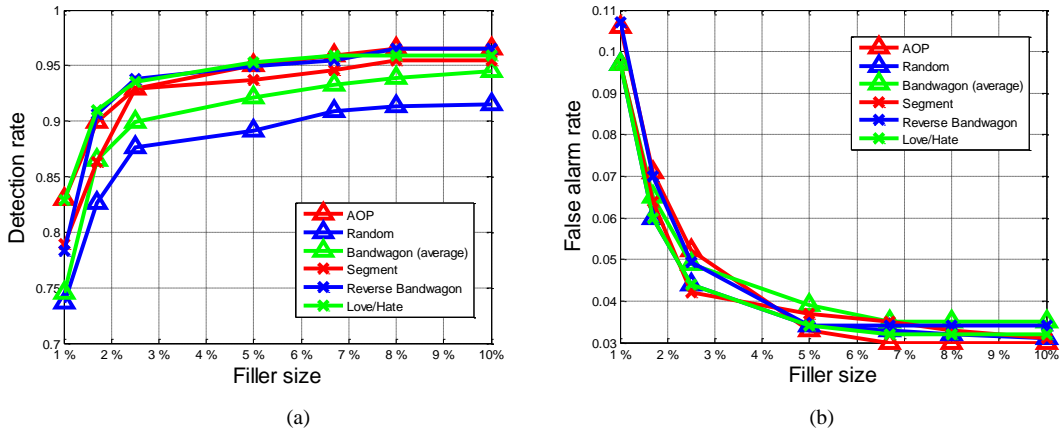

Fig. 6. The comparison of detection rate and false alarm rate in diverse attack models when attack size is 7.5% and filler size varies. (a) Detection rate. (b) False alarm rate.

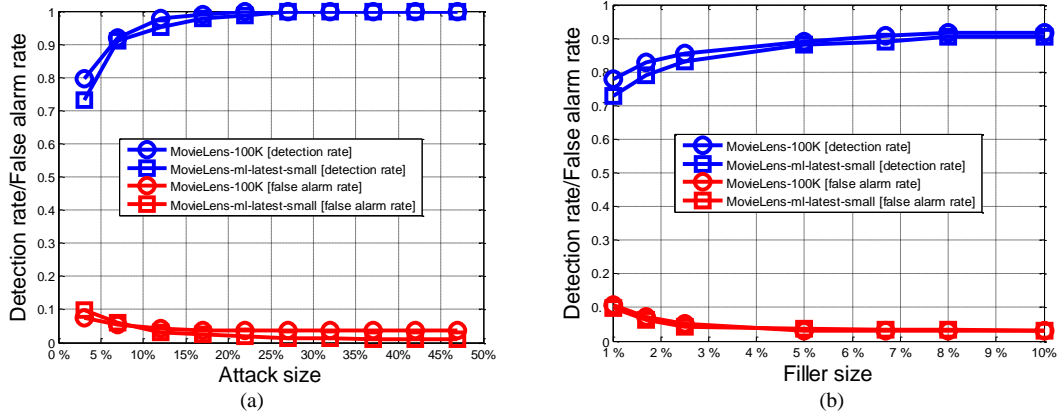

Fig. 7. The comparison of detection rate and false alarm rate in different datasets. All of them are nuke attacks. (a) The filler size is 5.2% and attack size varies in random attack. (b) The attack size is 7.5% and filler size varies in random attack.

To evaluate the detection performance of our detection method in different datasets, we conduct a list of experiments on both the MovieLens-100K and MovieLens-ml-latest-small datasets. As shown in Figure 7, the proposed method shows effective detection performance on both the two datasets regardless of the different filler sizes or attack sizes. Figure 7 (a) combines the detection rate and false alarm rate when the filler size is 5.2% and attack size varies. Figure 7 (b) combines the detection rate and the false alarm rate when attack size is 7.5% and filler size varies. We find that the detection rates increase very fast and achieve a stable level along with an increase in attack size or filler size. In Figure 7 (a), the detection rates reach close to 100% when attack size reaches 22.5%. the false alarm rates, on the other hand, consistently stay below 10% when the attack size gradually increasing. Similarly, in Figure 7 (b), the detection rates reach close to 90% when filler size reaches 5.2% and the false alarm rates also stay below 10% regardless of diverse filler sizes.

## 6. CONCLUSIONS AND FUTURE WORKS

"Shilling" attacks or "profile injection" attacks are serious threats to the collaborative filtering recommender systems (CFRSs). Since the existing features are unsuccessful to characterize the attack profiles and genuine profiles in some cases or published classification methods for detecting shilling attack are more rely on the existing features. In this paper, we provided a novel detection approach for detecting shilling attacks, which constructs a mapping model by exploiting the relationship between rating behavior and item distribution. Extensive experiments on both the MovieLens-100K and MovieLens-ml-latest-small datasets demonstrated the effectiveness of the proposed approach. To compare with the benchmarked method, our proposed method shows more optimistic detection performance. One of the limitations of our proposed method directly comes from new types of attacks incrementally, since they are generated over time in the context of real CFRSs. Besides, the detection performance of our method directly depends on the attack types and feature extraction in our training stage. In our future work, we intend to extend and improve attack detection in the following directions: 1) Considering more attack models such as Power user attack and Power item attack, etc.; 2) Developing a theoretical grounded method for the detection problem is also of interest. 2) Extracting more simpler and effective features from user profiles is also an open issue.

## ACKNOWLEDGEMENTS

The research presented in this paper is supported in part by the National Natural Science Foundation (61221063, U1301254), 863 High Tech Development Plan (2012AA011003) and 111 International Collaboration Program, of China.